\documentclass[12pt]{iopart}

\usepackage{iopams,graphicx}
\begin{document}

\title[Strangeness and Multi-Strangeness at LHC]
{(Multi)Strangeness Production in Pb+Pb collisions at LHC. 
HIJING/B\=B v2.0 predictions.}

\def\dmg{$^{1}$}
\def\cunew{$^{2}$}
\def\frank{$^{3}$}

\author{V.Topor Pop\dmg, J.Barrette\dmg, C.Gale\dmg, S.Jeon\dmg and
M.Gyulassy\cunew$^,$\frank}

\address{
\dmg \mbox{McGill University, Montreal, Canada, H3A 2T8}\\
\cunew \mbox{Physics Department Columbia University, New York, New York 10027, USA}\\
\frank \mbox{FIAS, J. W. Goethe Universitat, D-60438,Frankfurt am Main, Germany}\\
}


\begin{abstract}
Strangeness and multi-strangeness particles production
can be used to explore the initial transient field fluctuations 
in heavy ion collisions.  
We emphasize the role played by Junction anti-Junction (J\=J) loops and 
strong color electric fields (SCF) in these collisions. 
Transient field fluctuations of SCF  
on the baryon production in central (0-5 \%) Pb+Pb collisions at 
$\sqrt{s_{\rm NN}}$ = 5.5 TeV will be discussed  in the framework of
HIJING/B\=B v2.0 model, looking in particular to the predicted 
evolution of nuclear modification factors ($R_{\rm AA}$) 
from RHIC to LHC energies.  
Our results indicate the importance of a good description of the baseline
elementary $p+p$ collisions at this energy.
\end{abstract}



In previous publications \cite{prc75_top07} we studied 
the possible role of topological baryon junctions \cite{kharzeev96},
and the effects of strong color field (SCF)  
in nucleus-nucleus collisions at RHIC energies.
We have shown that the dynamics of the 
production process can deviate
considerably from that based on Schwinger-like estimates for 
homogeneous and constant color fields.
An increase of the string tension from $\kappa_0$= 1 GeV/fm, 
to {\em in medium mean values} 
of 1.5-2.0 GeV/fm and 2.0-3.0 GeV/fm, for d+Au and Au+Au 
respectively, results in a consistent description of the 
observed nuclear modification factors (NMF) $R_{\rm AA}$ 
in both reactions and point to the relevance of fluctuations 
on transient color fields. 
The model provides also an explanation of the baryon/meson anomaly,
and is an alternative dynamical description 
of the data to recombination models \cite{muller03}.

Strangeness enhancement \cite{rafelski_82}, 
strong baryon transport, 
and increase of intrinsic transverse momenta $k_T$ \cite{nuxu_04} 
are all expected consequences of SCF.
These are modeled  in our microscopic models as
an increase of the effective  string tension that controls the
quark-anti-quark ({\it q}$\bar{\it q}$) and 
diquark - anti-diquark (qq$\overline{\rm {qq}}$) pair creation rates
and the strangeness suppression factors.
A reduction of the strange ($s$) quark mass from $M_s$=350 MeV to the current
quark mass of approximately $m_s$=150 MeV, gives a
strangeness suppression factor $\gamma_s^1 \approx$ 0.70.
A similar value of $\gamma_s^1$ (0.69) is obtained by increasing   
the string tension from $\kappa_0$=1.0 GeV/fm to $\kappa$=3.0 GeV/fm
\cite{prc75_top07}.
Howeover, if we consider that Schwinger tunneling could explain
the thermal character of hadron spectra we can define an apparent 
temperature as function of the average value of string tension ($ <\kappa>$),
$T=\sqrt{3<\kappa>/4\pi}$ \cite{kharzeev_07}.
The predictions at LHC for initial energy density and temperature are 
$\epsilon_{\rm LHC} \approx$ 200 GeV/fm$^3$ and 
$T_{\rm LHC} \approx $ 500 MeV, respectively \cite{muller_06}.
Both values would lead in the framework of our model to an estimated increase 
of the average value of string tension 
to $\kappa \, \approx \,$ 5.0 GeV/fm at LHC energy.



\begin{figure}[h]
\vspace*{-0.2cm}
\begin{center}
\includegraphics*[width=12.0cm,height=6.0cm]{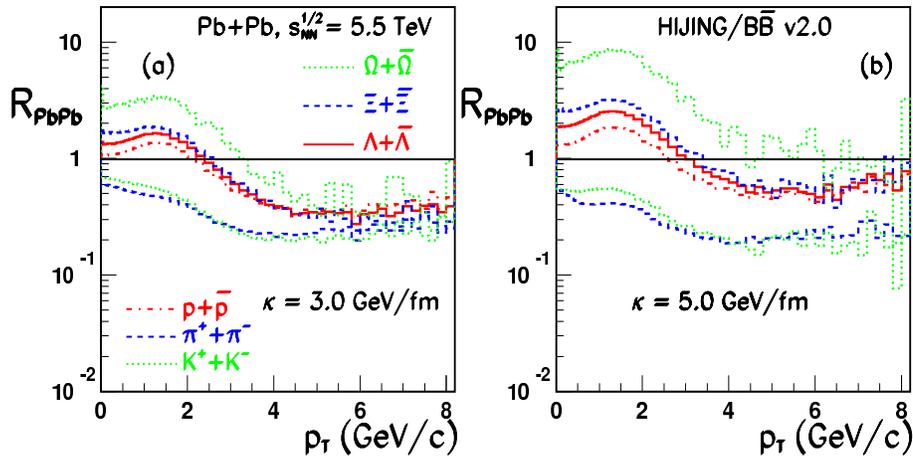}
\end{center}
\vspace*{-0.3cm}
\caption{HIJING/B\=B v2.0 predictions including SCF effects 
for NMF of identified particles. The results for proton and lambda particles
are for inclusive measurements.} 
\label{rpbpb}
\end{figure}

The {\it p}+{\it p} cross sections serve as a baseline
reference to calculate NMF for $A+A$ collisions ({\it R}$_{\rm {AA}}$).
In $p+p$ collisions high baryon/mesons ratio (i.e. close to unity) 
at intermediate $p_T$ were reported at $\sqrt{s_{\rm NN}}$= 1.8 TeV 
\cite{cdf_05}. These data could be fitted assuming a string tension 
$\kappa$=2.0 GeV/fm. This value is used in our calculations at  
$\sqrt{s_{\rm NN}}$= 5.5 TeV. This stresses the need for a reference 
$p+p$ measurements at LHC energies.

The predictions for NMF {\it R}$_{\rm {PbPb}}$
of identified particles at the LHC energy are presented in Fig.~\ref{rpbpb}
for two values of the string tension.
From our model we conclude that baryon/meson anomaly, will persist at the LHC 
with a slight increase for increasing strength of 
the chromoelectric field.
The NMF {\it R}$_{\rm {PbPb}}$ also 
exhibit an ordering with strangeness content at low and intermediate $p_T$.
The increase of the yield being higher for multi-strange hyperons 
than for (non)strange hyperons 
({\it R}$_{\rm {PbPb}}$($\Omega$) $>$ {\it R}$_{\rm {PbPb}}$($\Xi$)
$>$ {\it R}$_{\rm {PbPb}}$($\Lambda$) 
$>$ {\it R}$_{\rm {PbPb}}$($p$) ). 
At high $p_T > 4 GeV/c $ for $\kappa$=3.0 GeV/fm,   
a suppression independent of flavours is predicted due to quench effects.  
In contrast, this independence seems to happen 
at $p_T > $ 8 GeV/c for $\kappa$=5.0 GeV/fm.   

\begin{figure}[h]
\begin{center}
\vspace*{-0.3cm}
\includegraphics*[width=12.0cm,height=6.0cm]{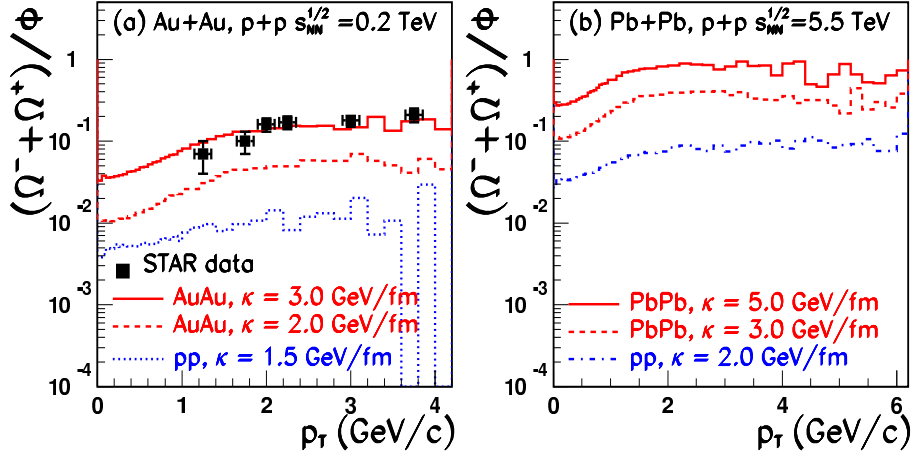}
\end{center}
\vspace*{-0.3cm}
\caption{Predictions of HIJING/B\=B v2.0 for  
the $(\Omega^++\Omega^-)/\Phi$ ratio as function of $p_T$ 
for RHIC (left panel) and LHC (right panel) energies. 
The experimental data are from STAR \cite{star_07}.}
\label{omphi}
\end{figure}

As expected, a higher sensitivity to SCF effects on the $p_T$ dependence 
of multi-strange particle yield ratio is predicted. As an example, 
 Fig.\ref{omphi} presents our results for the ratio 
($\Omega^{-} + \Omega^{+}$)/$ \Phi $ in central (0-5\%) Pb+Pb collisions
and  $p+p$ collisions.
The results and data at RHIC top energy are also included (left panel).

The mechanisms of (multi)strange particles production 
is very sensitive to the early phase of nuclear collisions, 
when fluctuation in 
the color field strength are highest. Their mid-rapidity yield favors 
a large value of the average string tension as shown at RHIC 
and we expect similar dynamical effects at LHC energy.
The precision of these predictions depens on our knowledge of  
initial conditions, parton distribution functions at low Bjorken-$x$,
the values of the scale parameter $p_0$, constituent 
and current (di)quark masses, energy loss for gluon and quark jets.

This work was partly supported by the Natural Sciences and 
Engineering Research Council of Canada and by the U. S. DOE 
under Contract No. DE-AC03-76SF00098 and
DE-FG02-93ER-40764. One of us (MG), gratefully acknowledges partial 
support also from FIAS and GSI, Germany.

\end{document}